\input psfig
\pssilent
\documentstyle[seceq]{ptptex}

\title{The close limit of colliding black holes: an update}

\author{Jorge Pullin}
\inst{Center for Gravitational Physics and Geometry,
Department of Physics, The Pennsylvania State University, 
104 Davey Lab, University Park, PA 16802, USA}

\recdate{
}

\abst{
This is the writeup of the talk I gave at the 
Yukawa International Symposium at Kyoto, Japan
on June 29, 1999. The talk summarizes the present
status of the close limit approximation for 
colliding black holes. 
}

\begin{document}

\maketitle

\section{Introduction}

\subsection{The three regimes of a black hole collision}

The collision of binary black holes is one of the primary expected
sources of gravitational waves to be detected by the broadband
interferometric gravitational wave telescopes currently under
construction, like the LIGO project in the US, the British/German GEO
project, the TAMA project in Japan and the French/Italian VIRGO project.

A collision of two binary black holes can be divided into three
distinct regimes.  Initially, the black holes spiral around each other
in quasi-Newtonian orbits.  The radius of the orbits decrease due to
the emission of gravitational radiation.  Let us call this period the
``inspiral" phase.  The gravitational waves produced during this phase
are well described by the post-Newtonian approximation.  Notice that
such an approximation does not provide a good description of the whole
spacetime, since it breaks down close to each hole (to first
approximation, the holes are singular point particles), but as long as
the holes are far apart, this is not expected to be relevant from the
point of view of the waveforms observed at
infinity\footnote{Technically, one can ignore the vicinity up to third
order post-Newtonian level \cite{Sch}.}.  The gravitational waves from
this phase of the collision correspond to a quasi-regular sinusoid
whose frequency and amplitude increases with time as the holes start
getting closer to each other, known as the ``chirp''.  Good
descriptions of this approximation applied to the binary black hole
case and appropriate references can be found in Blanchet et
al. \cite{Bl}.

When the separation of the holes is around 10 to 12 times the mass of
each individual holes, it is expected that the post-Newtonian
approximation breaks down.  It is not completely clear what is the
extent of the breakdown, since the post-Newtonian approximation leads
to an asymptotic perturbation series.  In fact, attempts are currently
being made to extend the validity of the domain of the approximation
using Pad\'e approximants \cite{pade}. In any event, there will be a
limiting separation of the holes such that if they are any closer, one
cannot use the post-Newtonian approximation. The domain that starts at
that point and continues up to the point in which the black holes form
a single black hole is expected to only be treatable by implementing
the evolution of the Einstein equations numerically. This has proven
to be a notoriously difficult problem. The state of the art of three
dimensional simulations of black hole collisions is such that at
present the codes can rarely evolve more than 30 or 40 in units of the
final black hole mass. One would need at least two orders of magnitude
more to be able to follow the black holes in the supposedly rapidly
decaying orbit below 10 $m$ in separation. Given that additional
resolution in 3D is very expensive, it is unlikely that the required
increase will be obtained simply by using more powerful computers; new
ideas appear to be needed.

Finally, when the black holes are close to each other, one can treat the
problem as a single distorted black hole that ``rings down'' into
equilibrium, evolving the distortions using perturbation theory.  This
is called the ``close limit approximation'' and 
will be the main subject of this talk.

\subsection{Why study the close limit?}
 
The study of  the final ringdown can be approached with three 
different perspectives. All of them have their own appeal, so I will
describe them in some detail:

a) {\em As a code check}. 
Whenever we finally have available a three
dimensional numerical code to integrate the Einstein equations for 
colliding holes, one could start the evolutions with the black holes
close to each other\footnote{As the experience in 
head-on collisions shows
\cite{etaletal} numerical codes can develop additional problems when
the black holes are close, let us ignore this detail here, since these
problems can usually be dealt with.}
The results should therefore coincide with those
of the close limit approximation. This point of view has actually been
pursued successfully in the head-on collisions. It turns out that 
even for this case the full numerical simulations have certain 
difficulties, and the close limit approximation can be used as a guiding
principle to build numerical codes \cite{BrAn}.

b) {\em To reach astrophysical conclusions}.  It is usually assumed
that the ringdown waveforms play no role in gravitational wave
detection.  This assumption is based on the fact that expectations are
that most black holes will occur in a mass range of a few solar
masses.  For such mass range, the ringdown occurs at too high a
frequency to be detectable by interferometric detectors, whereas the
inspiral phase sweeps the frequency range at which the detectors have
the peak sensitivity.  However, if the mass of the colliding holes is
higher, the inspiral's frequency becomes too low to be detected
whereas the ringdown is more easily detectable.  In fact, given that
larger masses also imply more radiated energy, these collisions become
easier to detect (for a detailed discussion see the papers by Flanagan
and Hughes \cite{FlHu}).  In fact, for an optimal mass range of about
300 Solar masses these collisions could even be visible by the initial
LIGO interferometers up to a distance of 200Mpc.  In fact, they are
likely to be the {\em only ones visible by the initial
interferometer}.  That such collisions might occur is not completely
out of the question, given our current ignorance about the population
of black holes.  Recent suggestions \cite{middleweight} that black
holes in the significant mass range might exist only reinforce this
possibility, although the existence of such holes is being currently
debated.  Even if one assumes that collisions like these take place,
the detectability of ringdowns is technically more involved than that
of the inspiral (essentially since many noises in the detector look
like ringdowns and also because template matching is hard since the
ringdowns are short lived in terms of number of cycles of
oscillation).  Jolien Creighton discusses these issues in detail
\cite{Creighton}.

The main drawback of using the ``close limit approach" in this context
is that one does not have the appropriate initial data to start the
problem. The initial data one would need to have ``astrophysically
meaningful" estimates of waveforms and radiated energies would
correspond to the endpoint of a black hole merger. But this is
precisely what we are unable to compute! The families of initial data
for colliding black holes usually considered are not supposed to be
physically realistic when one makes the separation parameters too
small.  This is essentially due to the fact that they are constructed
via ad-hoc superpositions based on mathematical convenience. If one
still insists on using them in the close regime (as we will do) one
has to admit that the results will not have a definite physical
justification.  Our experience after trying several families of
initial data is that the results (in terms of radiated energy) in the
end rarely differ by more than a factor of order unity.  Therefore
---at the level of an art-form more than that of a scientific
prediction---, one may be able to trust the results we present here
physically as order of magnitude estimates. This is the point of view
we will adopt from here on.

c) {\it To supplement numerical evolutions.} If the state of the art
of numerical relativity remains limited to few dozens of $m$ in terms
of the time length of the evolution, it would be useful to spend the
precious three dimensional evolution time of the codes ``coalescing"
the holes rather than following the ringdown of a single formed
hole. This approach has already been implemented in the case of
collapse of disks by Abrahams, Shapiro and Teukolsky \cite{AbShTe}.
This study used perturbations of Schwarzschild.  Currently under study
is the more general approach based on perturbations of Kerr for the
case of colliding holes by Baker, Campanelli and Lousto (the
``Lazarus/Zorro'' project at the Albert Einstein Institute in
Potsdam).

\subsection{Initial data}

To evolve a collision of black holes in the close limit one has to
start with a given family of initial data. As we mentioned in the
previous section, the physically correct initial data for close black
holes arising from an inspiral and merger is not available. The usual
families of initial data for binary black holes are obtained by ad-hoc
mathematical prescriptions. We will discuss in this section some of
the issues involved in such constructions. To have initial data for
general relativity means to have a three dimensional spatial metric
and an extrinsic curvature that solve the constraint equations of the
initial value problem of general relativity, i.e., the ``$G_{00}$''
and ``$G_{0i}$'' components of the Einstein equations.

A popular method of constructing solutions for these equations is the
Lichnerowicz-York conformal approach.  In this approach one assumes
that the three dimensional metric is conformally related to a given
fixed metric.  To simplify things, let us assume (as was done for
instance by Bowen and York \cite{BoYo}) that it is conformally flat.
If in addition one assumes that the trace of the extrinsic curvature
vanishes, the constraint equations simplify significantly. The
momentum constraint simply becomes the flat space divergence of a
tensor that is up to a factor the extrinsic curvature, and the
Hamiltonian constraint becomes an equation stating that the conformal
factor's Laplacian is related to the square of the extrinsic curvature
divided by the conformal factor to a given power.

The momentum constraint equations are easy to solve, and solutions
were introduced by Bowen and York \cite{BoYo}. In this approach the
tensor related to the extrinsic curvature completely determines the
ADM momentum and angular momentum of the slice. The solutions
constructed by Bowen and York depend on two vectors that coincide with 
the angular momentum and linear momentum of the slice.

\begin{figure}[t]
\hbox{\psfig{figure=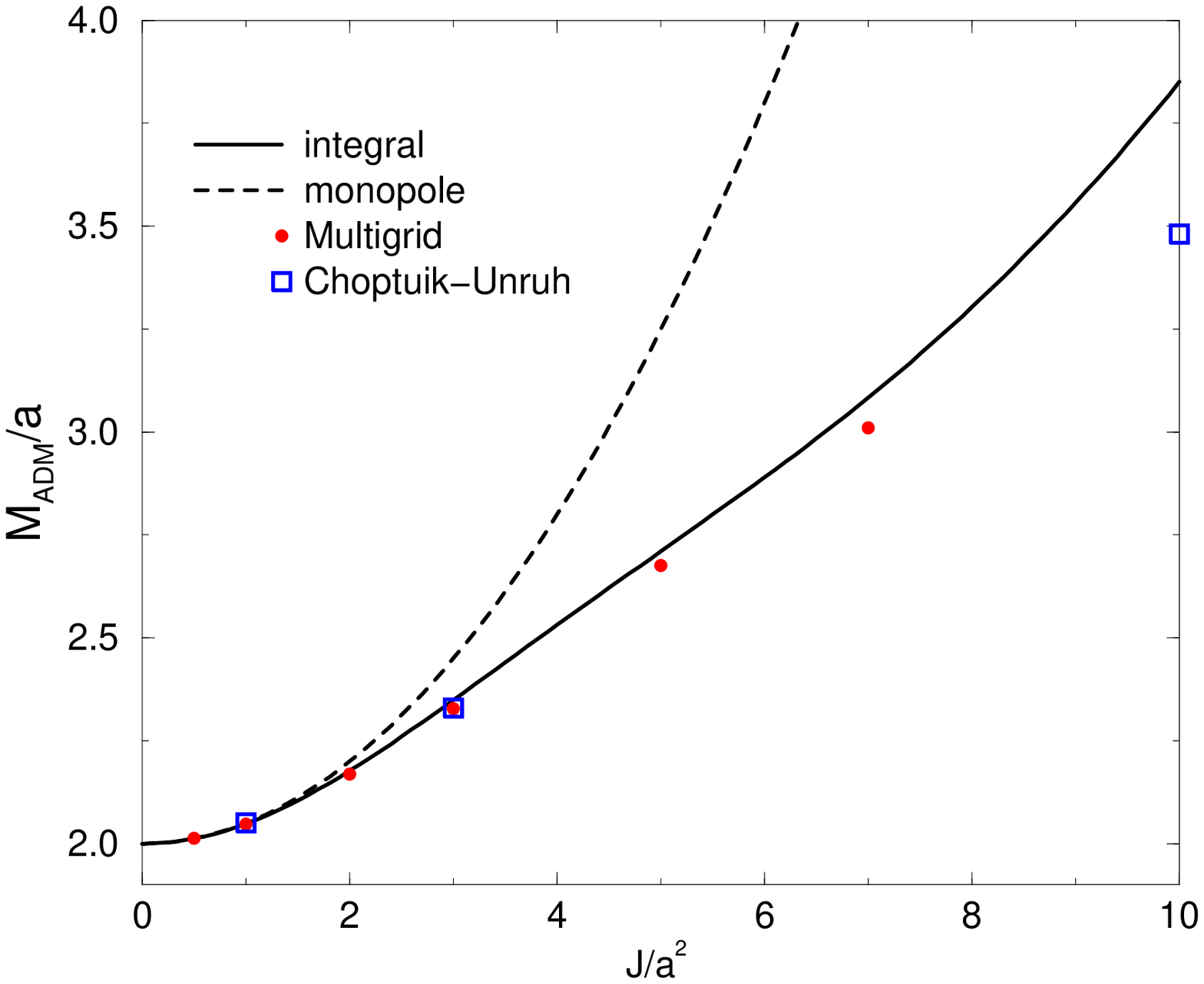,height=2.5in}
\psfig{figure=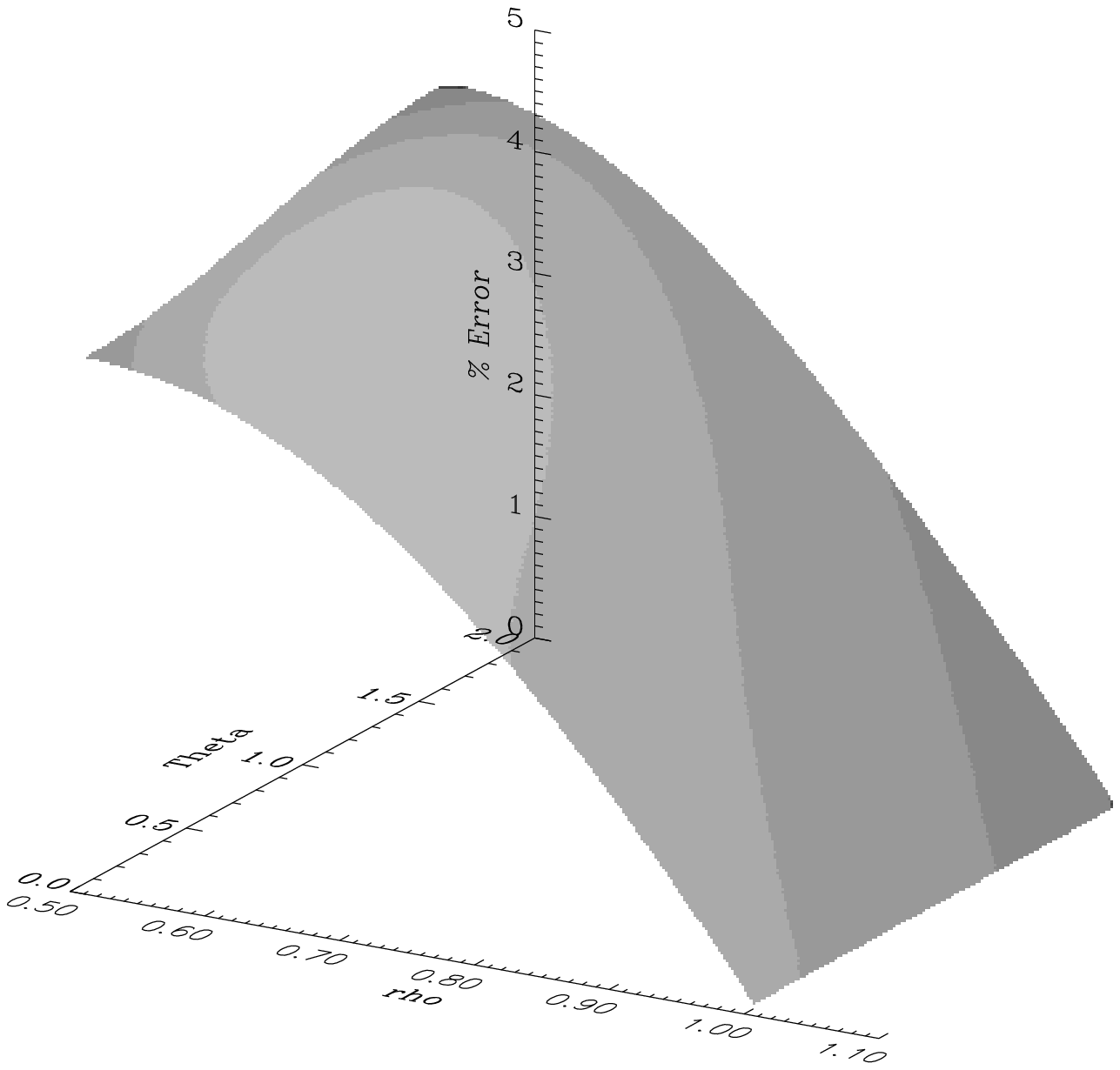,height=2.5in}}
\caption{Comparison of approximate solutions to the initial value
problem with a full numerical integration performed with a multi-grid
method, for the case of a single spinning Bowen--York black hole. The
figure at the left compares the ADM mass of approximate and numerical
solutions. The ``integral'' and ``monopole'' curves correspond to two
different ways of computing the mass, looking at the monopole term of
the conformal factor and using a Komar-type integral. The figure on
the right shows the percentile difference between the full numerical
solution for
$\psi$ (the fourth root of the conformal factor) 
and the second order approximation, for $J/a^2=4$,
as a function of $\rho,\theta$, where $a$ is the conformal radius of
the black hole ($M_{ADM}=2a+J^2/20a^3$ to second order in $J$). We see
that the approximation works very well even for moderately large
spins. }
\label{conformal}
\label{fig:phicompar}
\end{figure}

One then is supposed to solve the remaining nonlinear elliptic
equation for the conformal factor prescribing certain boundary
conditions.  This is usually achieved numerically, as discussed by
Cook \cite{Cook}.  Since the Bowen-York extrinsic curvatures are
linear in the momentum (linear or angular), for slow moving (or
rotating) holes, one can also seek approximate solutions for the
conformal factor expanding in powers of the momentum. To zeroth order
the solution simply corresponds to the vanishing of the Laplacian of
the conformal factor. This is the same equation one would have for a
time-symmetric situation ($K_{ab}=0$).  Since solutions to this case
with the topology of two holes are known (the Misner \cite{Mi} and
Brill-Lindquist \cite{BL} solutions), one immediately has approximate
solutions for moving holes, to zeroth order of approximation. It turns
out this is all we will really need for the close limit (in first
order perturbation theory). For higher orders in perturbation theory,
one can iterate the construction and explicitly obtain a solution for
the conformal factor as a power series in the momentum.  These
approximations work remarkably well, as shown in the figure
\ref{conformal} for the case of a single spinning hole
\cite{singlespin}.

\begin{figure}
\centerline{\psfig{figure=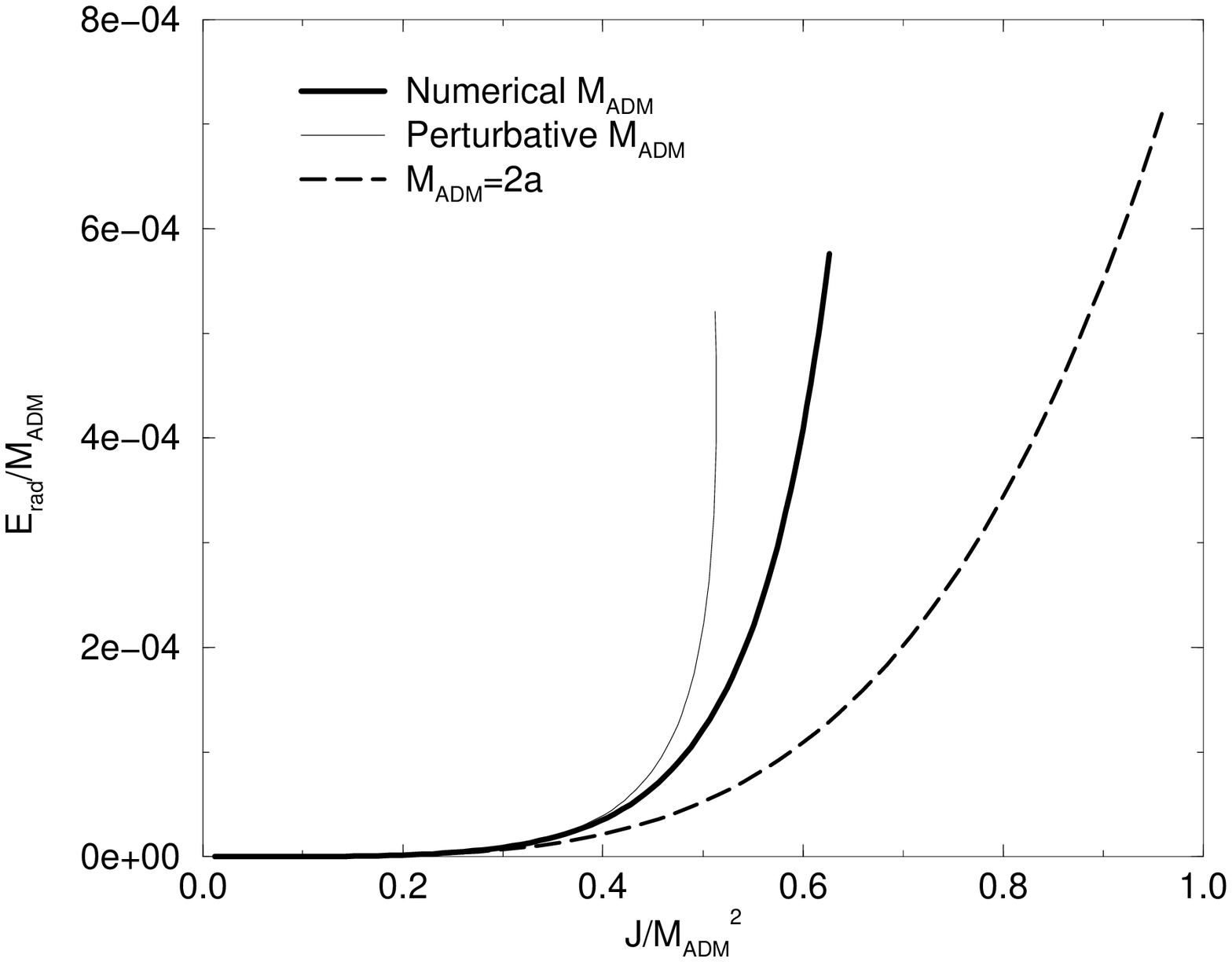,height=2.5in}}
\caption{The amount of energy radiated by a single spinning
Bowen--York black hole as it ``relaxes" to a Kerr black hole, computed
treating the spacetime as a perturbation of Schwarzschild (one can
view Kerr as a stationary perturbation of Schwarzschild, and the
radiated energy is given by the non-stationary part of the
perturbation). The amount becomes comparable to that produced by a
collision for values of the spin larger than 0.5 in terms of the Kerr
parameter $a$. The three curves correspond to different choices in how
to compute the ADM mass. As we argue in the text, the perturbative 
calculations of the mass are not very accurate, but one can recourse
to numerical calculations to get good estimates.}
\label{singlespin}
\end{figure}

An important drawback of the Bowen--York family of solutions is the
conformally flat nature of the spatial metric. This is especially
troublesome since neither a boosted Schwarzschild black hole nor a
spinning Kerr hole \cite{Ga} appear to admit slicings with spatially
flat sections. This means that the Bowen--York solutions will not
represent purely boosted or spinning black holes, but there will be
``additional radiation'', which in general will be larger the larger
the momenta of the holes. Since for realistic collisions one expects
the holes to be rapidly spinning, this is a serious impediment.  In
fact, one can consider \cite{singlespin} a single Bowen--York hole and
study its behavior treating it as a perturbation of a Schwarzschild
hole.  This has been done both for boosted and spinning holes, and as
shown in figure \ref{singlespin} for the spinning case, the total
radiated energy is low for small values of the spin. As we will see,
collisions of black holes rarely radiate more than $1\%$ of the holes'
mass, therefore one sees that the extra radiation in the Bowen--York
family is tolerable even for moderate values of the momenta.

Attempts have been made to generalize the Bowen--York ansatz 
to better
accommodate especially the spinning cases.  Krivan and Price
\cite{KrPr} and independently, Baker and Puzio \cite{BaPu}, have
proposed methods of solutions of the constraint equations.  The
Krivan--Price approach is based based on the fact that for solutions
that are ``conformally Kerr'' one can also find ways of superposing
holes.  Their solutions develop some undesired singularities, that in
the case of close black holes can be hidden by the common
horizon. These families of initial data have indeed been evolved
successfully in the close limit \cite{KrPr}.  The Baker--Puzio method
is based on choosing an ansatz for the spatial geometry that is able
of accommodating what one would intuitively consider the superposition
of the spatial metrics of two Kerr black holes, and then solving an
eikonal equation for the extrinsic curvature. This is a quite novel
approach in that one prescribes metric and solves for the extrinsic
curvature. The eikonal equations might however develop caustics and
other singularities, and the method has not completely been
implemented in practice. Both methods are up to present restricted to
axisymmetry, and therefore are not yet applicable for the most
interesting cases of inspiralling holes.  It appears that the only
solutions that one can construct that can reasonably accommodate
spinning holes in inspiralling situations will have to be built
numerically.

Other initial data proposals involve the use of Kerr-Schild ansatze
for the metric. They have both been pursued in the Cauchy
\cite{HuMaSh} and null \cite{Bishop} formulations. The close limit of
these families has not been explored yet.

\section{Evolution}

Once one has the initial data, one can proceed to evolve.  To achieve
this, the usual procedure has been to expand the initial data in terms
of an expansion parameter that goes to zero as the separation of the
holes goes to zero, and identify the radial coordinate in conformal
space with the radial coordinate in isotropic Schwarzschild
coordinates.  For cases involving boost or spin, one also keeps the
leading terms in $P$ or $S$, and assumes that $P$ and $S$ are of the
same order as the separation in conformal space $d$ in order to keep
mixed terms.  The first order departures from Schwarzschild are used
to evaluate the initial data for the Zerilli function, which is then
evolved using the Zerilli equation. It should be noticed that the
first order departures are of order $d^2$ in terms of the conformal
separation for non-moving holes and also have terms of order $P d$ for
boosted holes. Here we face an inevitable difficulty, which in the end
becomes problematic, at least for inspiralling holes. When one
considers collisions of black holes with arbitrary boosts and spins,
the problem is really multiparametric, and one is really pushing
things by insisting on fitting the problem into the usual framework of
black hole perturbation theory, where one starts assuming a
one-parameter family of space-times.

To evolve first order perturbations of black holes one has available
several formalisms. They are all equivalent, but the details are
significantly different. Let me concentrate on two of the most popular
approaches. One of them is based on the Newman-Penrose formulation and
leads (in the case in which the background space-time corresponds to a
rotating black hole) to the so-called Teukolsky equation. This
equation is a (complex) equation for the linearized part of one of the
components of the Weyl spinor. In the case of a non-rotating
background the Teukolsky equation reduces to the so-called
Bardeen--Press equation. For a variety of historical reasons we have
not used these formalisms in our approach, but rather used a different
formalism which we broadly call Regge--Wheeler--Zerilli (RWZ)
formalism. This formalism was constructed by treating separately the
even and odd parity portions of the linearized perturbations. In both
cases a real function is constructed out of the linearized components
of the metric and satisfies a linear equation. For the case of
even-parity perturbations the equation is called the Zerilli equation
and for the odd-parity perturbations it is called the Regge--Wheeler
equation. They are both equations for a single real function enconding
the relevant gravitational degree of freedom.  In the even parity
case, the so-called ``Zerilli function'' is constructed with the
components of the perturbative metric and its first time
derivatives. Formulas for its construction are available that are
invariant under first order coordinate transformations (gauge
transformations) \cite{CPM}. Similar formulations are available for
the odd-parity perturbations.

For the case of head-on collisions of momentarily stationary and
boosted black holes, as well as for non-head-on collisions of
non-spinning holes, the first order perturbations are only
even-parity, and therefore the whole problem can be treated solely
with the Zerilli equation. This was in part the historical reason for
looking at this formalism, since it is somewhat simpler than the
Teukolsky one for these initially important cases, and yet
applicable. At first order in perturbation theory, the Zerilli and
Regge-Wheeler formalisms use functions that involve only first order
time derivatives of the initial metric. The Bardeen--Press approach
requires one further derivative, which means that one has to use the
Einstein equations in addition to the initial value equations to
construct the initial data. This is a bit more cumbersome, and in
fact, can introduce differences \cite{Lousto} 
depending on how one keeps orders in 
solving the Einstein equations, so it should be kept in mind
in comparing the formalisms.

\begin{figure}
\hskip 3cm \psfig{figure=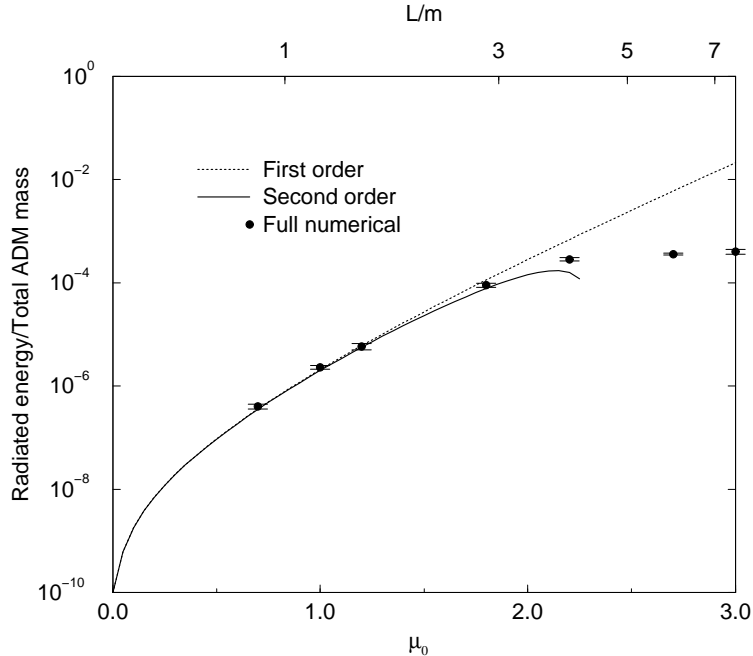,height=3.5in}
\caption{The radiated energy in a collision of two
momentarily-stationary black holes (the Misner problem) compared with
the results of full numerical simulations of the NCSA/Potsdam/WashU
group. We see that the approximation works well for black holes that
are closer than about six times the mass of each hole.}
\label{misner}
\end{figure}

Figure \ref{misner} shows the radiated energy in a collision of two
momentarily stationary black holes (Misner problem), 
as a function of the initial separation. We see here that the close
approximation predicts very well the radiated energy up to separations
of about six times the mass of each individual hole, when compared
with full numerical simulations.

The figure also includes results for second order perturbation
theory. The formalism is described in detail in
\cite{GlNiPrPuphysrep}. What is clear from the figure is that the
perturbative formalism is self-consistent, that is, it is able to
predict via recourse to higher order perturbations when it will fail
to agree with the numerical results. The results are even more impressive for waveforms as shown in the next figure. 
\begin{figure}
\hskip 3cm \psfig{figure=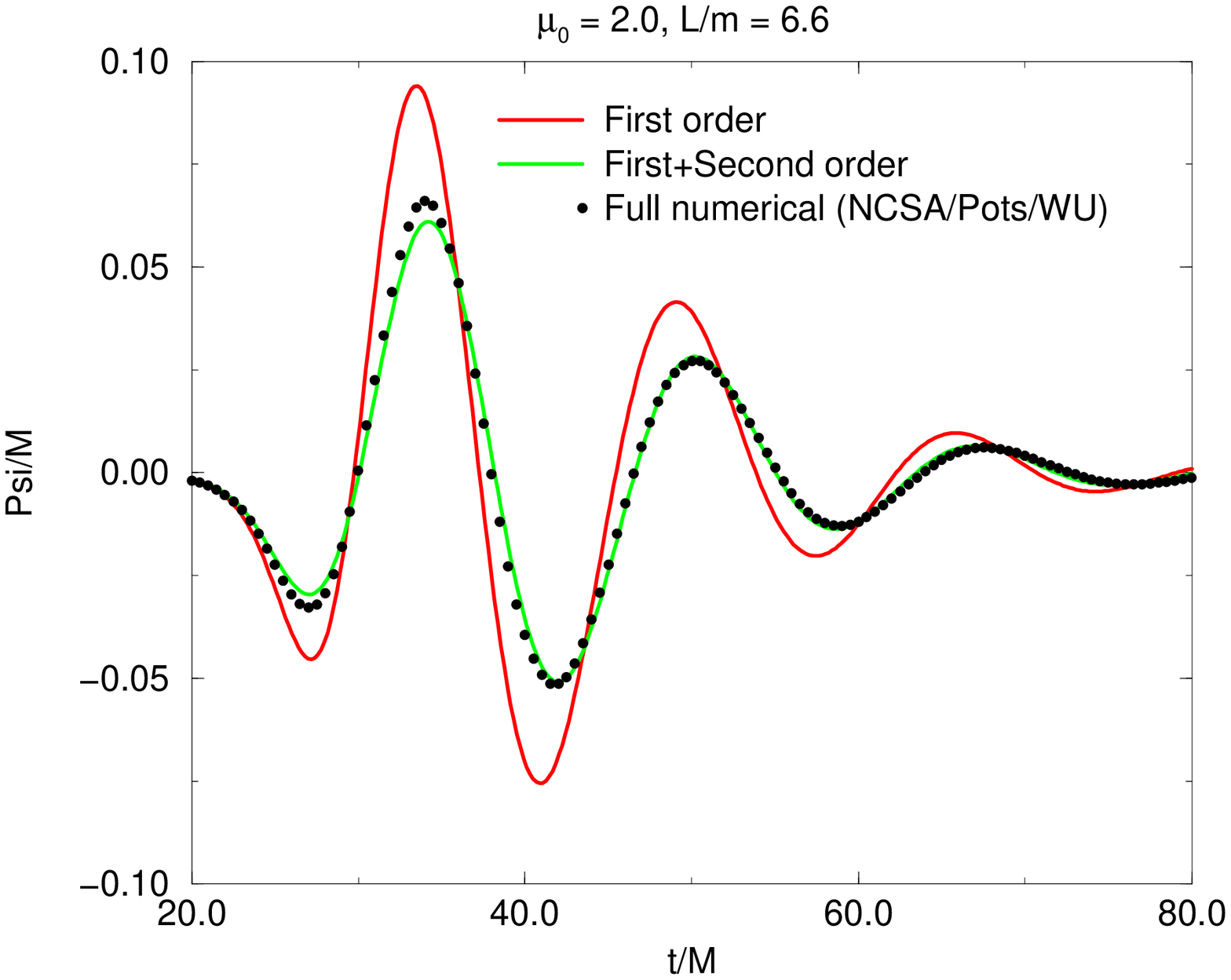,height=3.5in}
\caption{First and second order waveforms. Because the first and second
order Zerilli functions are not the coefficients of an expansion of 
a function, it makes no sense to compare them. We therefore present the
time derivative of the first order Zerilli function and a second order
correction to it. These expressions squared are proportional to the 
radiated power, and convey information about the gravitational
waveform.}
\end{figure}
This figure corresponds to a region of parameters in which the second
order correction is maximal, that is, just before perturbation theory
breaks down. It should be emphasized that the numerical results have
certain uncertainties as well, as discussed in \cite{BrAn}, so the
agreement is probably even better than depicted.

\begin{figure}[h]
  \centerline{\psfig{figure=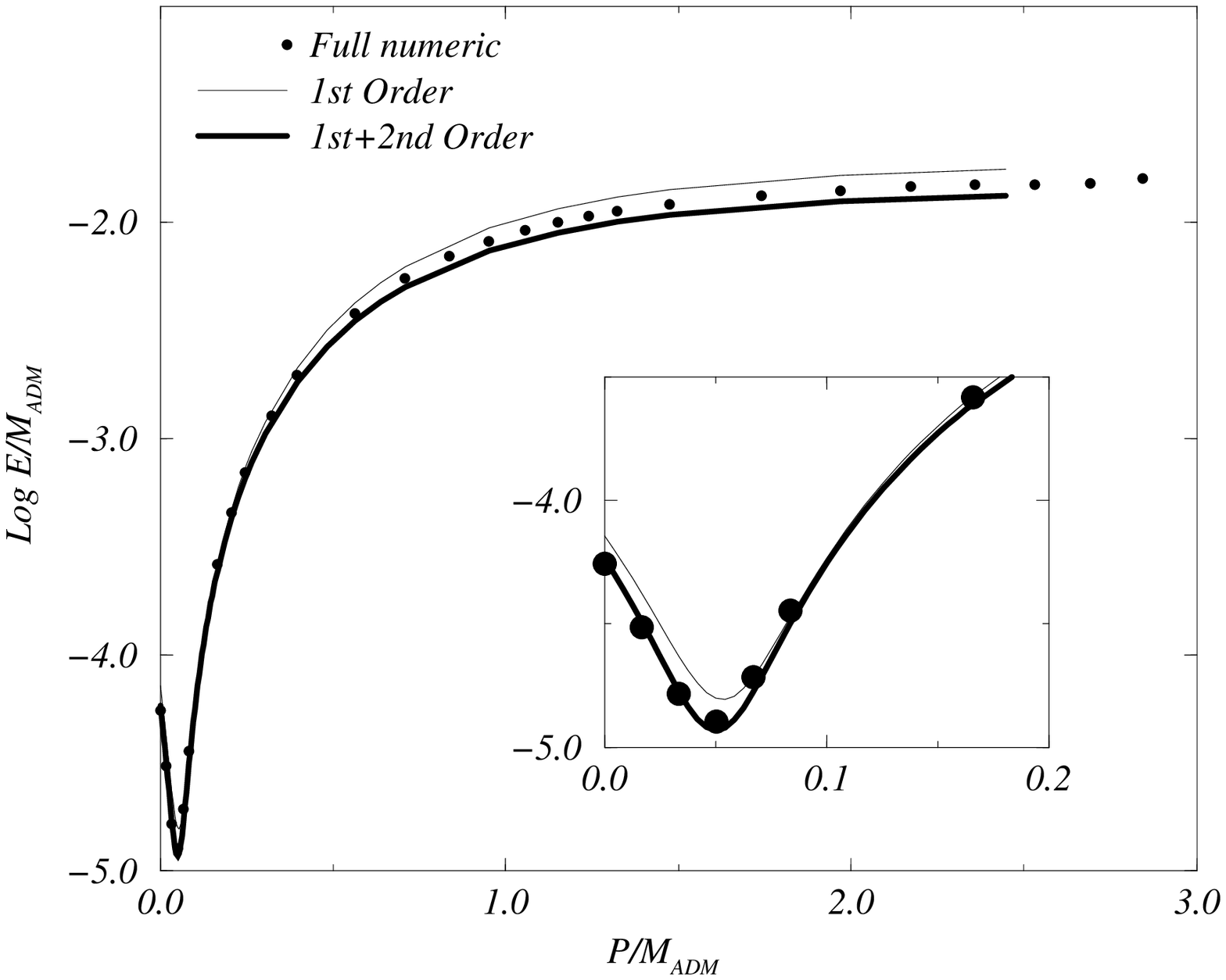,height=3.5in}}
\caption{Radiated energy in head-on black hole collisions as function
of the momentum for a separation of $\mu_0=1.5$, $L_{\rm phys}/(0.5
M_{ADM})= 5.5$.  Depicted are the close-slow approximation and the
full numerical results of the Potsdam/NCSA/WashU group.  Even for
large values of the momentum, the first order results overshoot and
the first plus second order undershoot the numerical results by only
$20\%$. The inset shows the ``dip'' region.}
\label{boostfig}
\end{figure}

These results are illustrative of what one can achieve in the head-on
collision case with the close limit. One can also discuss collisions
of boosted black holes \cite{boost1,GlNiPrPuboost} to first and second
order in perturbation theory, and the agreement with the numerical
results are even more attractive. In figure \ref{boostfig} we show
the agreement with numerical results for the energy.
We see two remarkable things in the energy plot: first of all, the
approximation works very well for large values of the
momentum. Remember we argued initially that we would be doing a
``slow'' approximation, in the fact that we ignored the
right-hand-side of the Hamiltonian constraint. Why is it then that the
energies keep on agreeing well for large values of $P$? The answer
lies in the structure of the equations of the initial value
problem. The equation satisfied by the extrinsic curvature is
linear. Therefore as we increase the momentum, the extrinsic curvature
grows without bound. In fact it grows linearly. The Hamiltonian
constraint, however, due to its non-linear structure, implies that the
conformal factor has a weak dependence on the momentum. As a
consequence, for large values of the momentum, the initial data is
completely ``dominated'' by the extrinsic curvature piece. We are
doing a very poor job of accounting for the conformal factor with the
``slow'' approximation, but since the conformal factor is in norm very
small with respect to the extrinsic curvature, the evolution is
completely dominated by the extrinsic curvature, for which we have an
exact solution! One should be a bit cautious with this statement on
one occasion: in the calculation of the ADM mass. The ADM mass is {\em
completely} dominated by the conformal factor, therefore it is very
poorly approximated by our technique. But the ADM mass can be computed
numerically in the initial slice, and since the collisions radiate a
comparatively small amount, the approximation is good throughout the
evolution.

The second interesting aspect of the energy plot for the boosted
collision is given by the ``dip'' in the energy that occurs as one
increases the momentum. This is related to the previous point. If one
starts the plot from the left, initially there is zero momentum, so
one is simply recovering the results of the Misner case. In such case,
all the radiation is produced by the conformal factor since the
extrinsic curvature vanishes identically. As one increases the
momentum, the portion of the initial data coming from the conformal
factor and that of the extrinsic curvature ``compete'' with each
other, and actually cancel each other, giving rise to the dip. As the
momentum is increased further, the extrinsic curvature dominates. The
cancellation at the dip implies that first order perturbation theory
actually does not work too well, in spite of the fact that nominally
we are in the optimal regime of applicability. Since there is a
cancellation occurring one needs higher orders to account properly for
things, as the energy plot shows.

\section{Collisions with net angular momentum}

The most interesting collisions of black holes are not the head-on
ones of course but the ones with angular momentum. In such case the
immediate reaction is to think that the space-time should be
approximated as a perturbation of a Kerr black hole. We shall see that
this is not necessarily the case. There are several reasons why it
might not be better to consider perturbations of Kerr.

To begin with, the whole perturbative paradigm consists in assuming
one has a background metric and then ``small departures''
characterized by a dimensionless parameter $\epsilon$. Consider the
collision of two non-spinning holes. In the ``close limit''
approximation the way we have set up things is to assume that both the
separation of the holes $d$ and their linear momenta $P$ are small. As
a consequence the total angular momentum of the holes $L=P d$ will be
small. In the ``close limit'' the angular momentum goes to zero. That
is, when we make the perturbative parameter small in such a family of
initial data, we recover the Schwarzschild spacetime and not the Kerr
spacetime. We shall in fact see that for this case (non-spinning
holes) one is indeed better off not using Kerr perturbations in
practice. Moreover, if one insists in using Kerr perturbations, the
perturbative formalism one sets up is at best peculiar. This is due to
the fact that {\em the perturbative parameter} (essentially the
angular momentum) {\em now appears in the background spacetime}. And
to all orders in perturbation theory. This is not the usual way
perturbation theory is set up. We have carried out calculations of
this sort but we will shortly see that these conceptual difficulties
eventually lead to confusions.

What if the holes are spinning? In such a situation one would
presumably be better off considering the problem as a perturbation of
a Kerr spacetime, but there are caveats. Is one going to consider the
spins as fixed and determining the background and then use linear
momenta and separation as ``small'' and ``comparable'' perturbative
parameters? One might, but it would be odd in the sense that the
angular momentum of the system should be taken into account when
computing the total angular momentum. However, the orbital angular
momentum is a significant component of the total angular momentum, we
are back at the same problem as before: the background will depend
(maybe more mildly) on the perturbative parameter. To add to the
difficulties, the families of Bowen and York do not represent Kerr
black holes well individually, so if one is interested in studying
situation with high spins in the individual holes, one will be adding
a lot of spurious radiation.

One might be facing an unsolvable problem in the sense that the
Schwarzschild solution has a more ``robust'' nature than the Kerr
solution. That is, all concentrations of energy that are roughly
spherical are close to the Schwarzschild solution outside. Rotating
configurations only have exterior Kerr fields if there is a precisely
tuned set of multipoles in the field. Two distinct concentrations of
energy, like black holes, that inspiral towards each other might
simply  not look from the outside like a single rotating black
holes, multipole-wise.

In the end perhaps the best way of sorting out these issues is to
attempt to apply the perturbative formalism for these problems, and
see what is the outcome. One can be conservative: in parameter regions
where all formalisms agree, one can be quite confident in the results,
and discard other results until confirmed in other ways. We are in the
process of doing so. Currently we have only completed the non-head-on
collision of two non-spinning Bowen--York holes \cite{science}. We
have evolved it with both the Zerilli and the Teukolsky formalism.  To
achieve the evolutions with the Teukolsky formalism we needed several
intermediate results. To begin with, there was virtually no experience
with the Teukolsky equation in the time domain, largely because it is
a $2+1$-dimensional problem. Krivan, Laguna, Papadopoulos and
Andersson have now written a code
\cite{KrLaPaAn} that integrates the Teukolsky equation in the time
domain. That is the code we are using for evolution. Given the lack of
experience with the Teukolsky equation in the time domain, we had to
set up formulas relating the metric and extrinsic curvature to the
initial data for the Teukolsky function. This is somewhat complicated
technically, but it can be achieved \cite{CaLoBaKhPu}.

\begin{figure}
\centerline{\psfig{figure=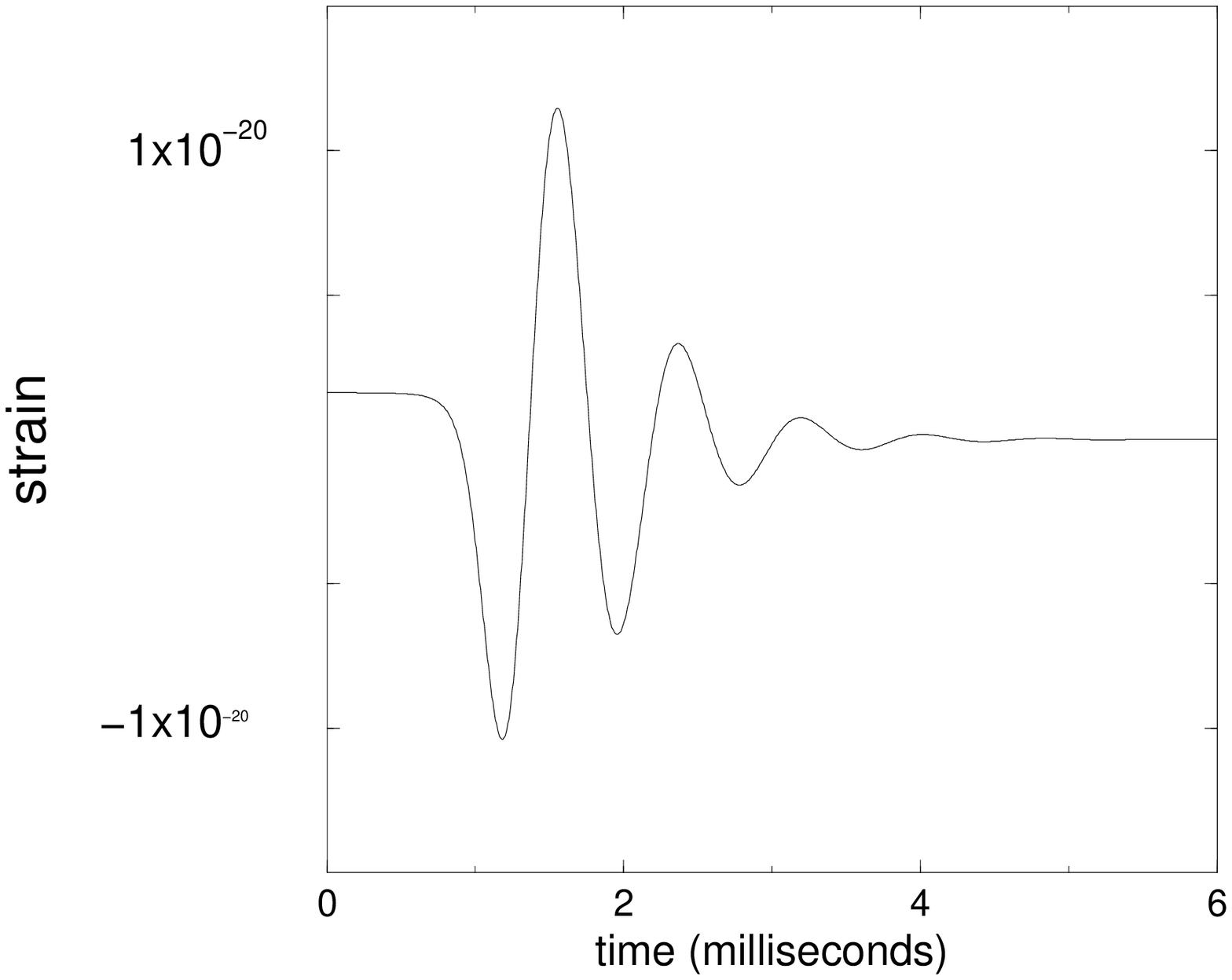,height=60mm}
\psfig{figure=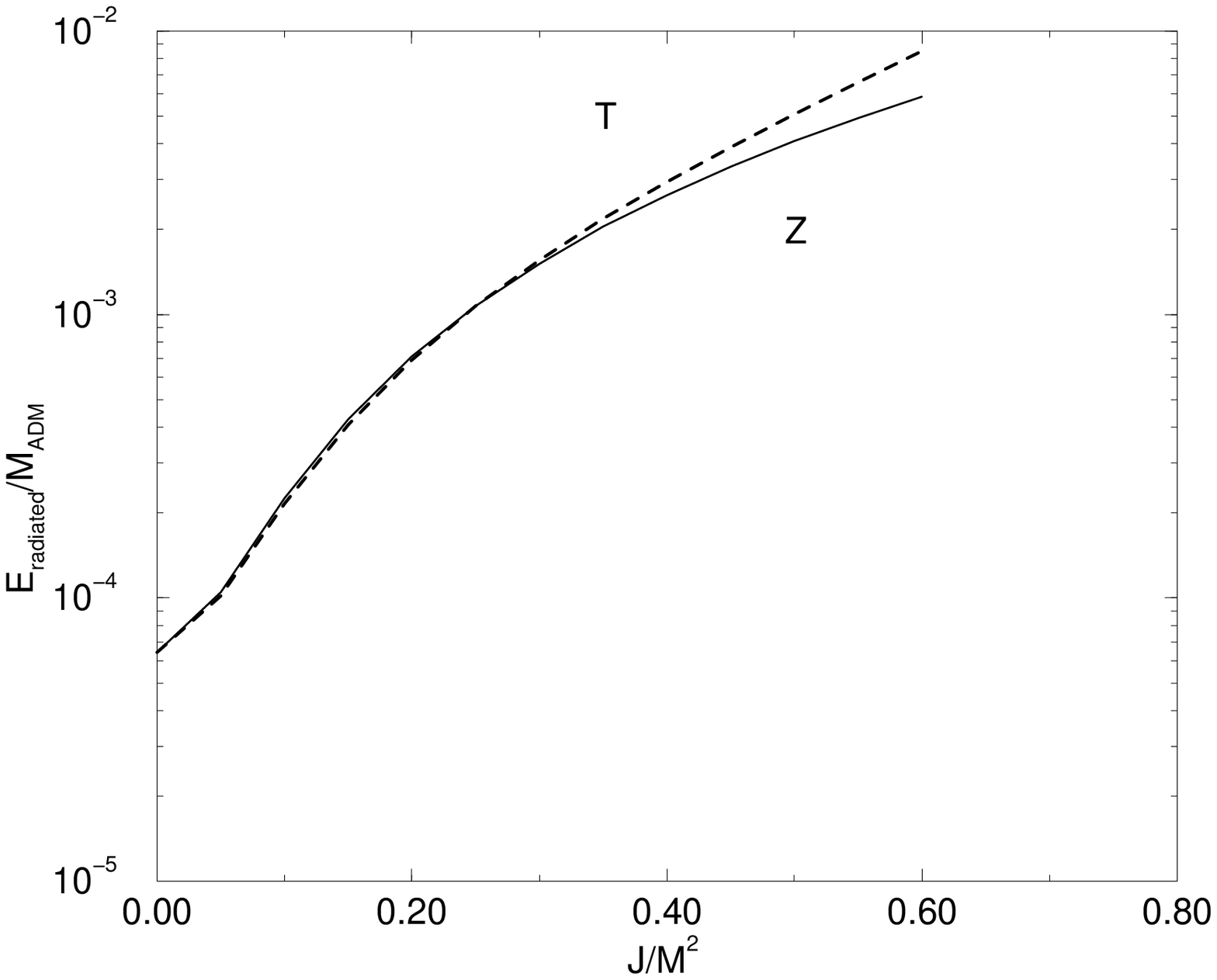,height=60mm}}
\caption{ The figure on the left shows the strain amplitude in the
equatorial plane as a function of time, producted by a $10M_{\odot}$
black hole binary going through its ringdown phase at a distance of
100\,Mpc from the detector.
We assume that the
detector is oriented for maximum sensitivity to the radiation in the
orbital plane. (For an L shaped laser interferometer, one arm
of the L would have to have the orientation just described for
the bar; the other arm would have to be perpendicular to the orbital
plane.)
On the right is shown the fraction
of the mass of the system radiated as gravitational waves as a
function of the normalized initial angular momentum of the collision.
}
\label{inspiral}
\end{figure}

Figure (\ref{inspiral}) depicts the waveforms and energy radiated for 
the non-head-on collision of two black holes.
The result shown is for two black holes initially separated in
conformal flat space by $d=1.8$ in terms of the mass of each hole. (If
one were considering a Misner type geometry, the proper separation
measured along the geodesic threading the throats would be $5.5$ in in
the same units \cite{etaletal}.)  The curve labeled $Z$ shows results
for linearized perturbation calculations using the Zerilli equation;
the curve $T$ shows the result of ``hybrid perturbation'' calculations
using the Teukolsky equation.  The two results diverge around
parameter values of $J/M^2=0.4$ to $0.5$, and this is a reasonable
limit to take for the applicability of perturbation estimates. We note
that the Teukolsky results lie above the Zerilli results, and this
weakly suggests that the Zerilli-based estimates are more
accurate. (In close limit estimates for head-on collisions, linearized
results always overestimated the nonlinear --- i.e., numerical
relativity --- results.)

Summarizing, we see that in the close limit the collisions do not seem
to radiate more than $1\%$ of the mass of the holes. This limitation
is robust, in the sense that we already saw it in the boosted head-on
collisions: if one attempts to increase the radiation by boosting the
black holes harder, one also increases the initial ADM mass and
therefore the radiated fraction of the energy in the end does not
increase. The one percent figure is smaller than numbers that have been
traditionally assumed for data analysis purposes \cite{FlHu}.

Having at hand collisions without axisymmetry, one can ask questions
about the radiation of angular momentum. A priori these are very
interesting questions since it is expected that black holes will
inspiral towards each other with too much angular momentum, in the
sense of possessing more angular momentum than that needed to make the
final resulting black hole extremal. Presumably this excess angular
momentum has to be radiated somehow. It is not expected that this
would happen in the final instants of the collision, but nevertheless
it would be instructive to see what happens in these final moments. 

We have computed the radiated angular momentum in both the Zerilli and
Teukolsky formalisms. At the moment however, it is not clear if these
calculations are appropriate. We find that the radiated angular
momentum disagrees in both calculations. We have eliminated all
possible sources of error by simply evolving the same initial data
with the Teukolsky and Zerilli codes and checking that if one
eliminates from the Teukolsky evolution equation (but not from the
initial data) the angular momentum dependent terms, the results agree
with those of the Zerilli evolutions.  It appears that the addition of
those (inconsistent perturbatively, as we argued above) small terms
changes the predictions dramatically. This is not entirely
surprising. The radiated angular momentum is a more subtle quantity to
compute than the energy (where both formalisms agree quite well). The
latter is basically a sum of squares, whereas the angular momentum is
given by a correlation of modes. A small phase shift in one of the
modes will therefore have no impact whatsoever on the calculation of
the energy, but would change dramatically the angular momentum
radiated. Apparently this is the effect of the extra terms in the
Teukolsky equation, and therefore the calculation in this formalism
predicts much more radiated angular momentum.

There clearly is more to be understood in the comparison of Teukolsky
and Zerilli calculations for the close limit of colliding black
holes. This will require working both formalisms to higher
order. Progress in setting up a second order formalism for the
Teukolsky equation is being made by Campanelli and Lousto \cite{CaLo}.

\section{Summary}

The close limit of black hole collisions has taught us several things
about black hole collisions. The formalism is not capable of
addressing the most interesting questions in the subject, but it
allows us to tackle certain issues in a degree of concreteness that
the full numerical simulations are currently lacking. Further work is
needed to complete the understanding of the close limit of
inspiralling black holes with spin. The whole subject has spawned
interest in the initial data problem and progress is being made on
this front too. The application of perturbative techniques to extend
the life of full numerical codes also opens a new avenue for synergy
between numerical and analytical work. In my opinion this synergy will
be vital to allow the final tackling of the problem of two colliding
black holes.

\section{Acknowledgments} I wish to thank the organizers 
for the invitation to
this wonderful conference.  This paper summarizes work done with many
collaborators: among the ones who have contributed the most are John
Baker, Reinaldo Gleiser, Gaurav Khanna, Pablo Laguna, Hans-Peter
Nollert, Richard Price. I am also grateful to Pete Anninos, Andrew
Abrahams, Greg Cook, Steve Brandt and Ed Seidel for help with
comparisons with full numerical results. I am indebted to many other
people for insights and discussions.  This work was supported in part
by the National Science Foundation under grants NSF-INT-9512894,
NSF-PHY-9423950, NSF-PHY-9407194, research funds of the Pennsylvania
State University, the Eberly Family research fund at PSU.  JP
acknowledges support of the Alfred P. Sloan and John Simon Guggenheim
foundations. I wish to
thank the Institute for Theoretical Physics of the University of
California at Santa Barbara for hospitality during the completion of
this work.

\end{document}